\documentclass[
reprint,
aps,
]{revtex4-2}

\usepackage{booktabs}
\usepackage{placeins} 
\usepackage{url}
\usepackage[hyperindex,breaklinks]{hyperref}
\hypersetup{linkcolor=blue, citecolor=red, colorlinks=true}

\usepackage{capt-of}
\usepackage{graphicx}
\usepackage{dcolumn}
\usepackage{bm}

\usepackage{amsmath, amssymb, physics}

\newcommand{\ii}{\mathrm{i}} 

\begin{document} 

\title{
Directionality and quantum backfire in continuous-time quantum walks from delocalized states: Exact results
}

\author{Jefferson J. Ximenes$^{1}$}
\thanks{jeffersonximenes@ufu.br}

\author{Marcelo A. Pires$^{2}$}
\thanks{piresma@cbpf.br}

\author{José M. Villas-Bôas$^{1}$}
\thanks{boas@ufu.br}

\affiliation{
$^{1}$Instituto de Física, Universidade Federal de Uberlândia, 38400-902 Uberlândia-MG, Brazil
\\ 
$^{2}$Centro Brasileiro de Pesquisas F\'{\i}sicas, Rio de Janeiro - RJ, 22290-180, Brazil
}

\begin{abstract}
We derive analytical results for continuous-time quantum walks from a class of initial states with tunable delocalization. The dynamics are governed by a Hamiltonian with complex hopping amplitudes. We provide closed-form equations for key observables, revealing three notable findings: 
(1) the emergence of directed quantum transport from completely unbiased initial conditions; 
(2) a quantum backfire effect, where greater initial delocalization enhances short-time spreading but counterintuitively induces a comparatively smaller long-time spreading after a crossing time $t_{\mathrm{cross}}$; and
(3) an exact characterization of survival probability, showing that the transition to an enhanced $t^{-3}$ decay is a fine-tuned effect. Our work establishes a comprehensive framework for controlling quantum transport through the interplay between intermediate initial delocalization and Hamiltonian phase.
\end{abstract}

              
\maketitle

\paragraph{\textbf{Introduction.}}

\textcolor{black}{Quantum walks (QWs) serve as versatile platforms for a wide range of fields, spanning from quantum algorithms~\cite{portugal2013quantum,kadian2021quantum,mulken2011continuous,khalilipour2025review,desdentado2025quantum} to the simulation of physical phenomena such as topological phases~\cite{wu2019topological}, 
relativistic effects~\cite{strauch2006relativistic},
two-dimensional transport phenomena~\cite{tang2018experimental}, and laser physics~\cite{heckelmann2023quantum}. Their applicability extends even to interdisciplinary areas such as finance~\cite{de2025potential} and 
complex networks~\cite{biamonte2019complex}.}

The framework of QWs encompasses both discrete-time (DTQWs~\cite{aharonov1993quantum}) and continuous-time (CTQWs~\cite{farhi1998firstCTQW}) variants. 
Conventional versions of \textcolor{black}{these models} exhibit remarkable properties such as ballistic spread and bimodal distributions, but a richer phenomenology emerges \textcolor{black}{in nonstandard scenarios}~\cite{pires2019multiple,pires2020quantum,soni2025advances,qiang2024quantum,zhou2021review,xia2019random}.
\textcolor{black}{A key degree of freedom in these systems is the initial condition: QWs can be initiated from} local states or extended (also called nonlocal or delocalized) states~\cite{abal2006quantum,abal2006effects,strauch2006relativistic,strauch2006connecting,de2010tailoring,annabestani2010asymptotic,machida2013quantum,zhang2016creating,orthey2017asymptotic,orthey2019connecting,orthey2019weak,ghizoni2019trojan,martin2020optimizing,vieira2021quantum,naves2022enhancing,buarque2022rogue,engster2024high,su2019experimental,chaves2023transport}.

\textcolor{black}{First studies on nonlocal initial conditions appeared in 2006~\cite{abal2006quantum,abal2006effects,strauch2006relativistic,strauch2006connecting}, revealing that delocalization can enhance entanglement between internal (spin) and external (position) degrees of freedom~\cite{abal2006quantum} and modify the scaling of survival probabilities~\cite{abal2006effects}. Subsequent research demonstrated that extended initial states allow for the tailoring of wavepacket propagation~\cite{de2010tailoring}, the generation of non-classical cat states~\cite{zhang2016creating}, and the study of rogue waves~\cite{buarque2022rogue}. Furthermore, delocalized states have been identified as effective resources for state transfer protocols~\cite{engster2024high}, and their experimental feasibility has been successfully demonstrated using initial superposition states~\cite{su2019experimental}.}

Recently, Ref.~\cite{chaves2023transport} investigated CTQWs initiated from both fully localized ($|0\rangle$) and fully delocalized ($|\pm x_0\rangle$) states. While these extreme cases have been thoroughly examined, the intermediate regime remains largely unexplored. Our work addresses this gap.


\paragraph{\textbf{Model.}}

We analyze a continuous-time quantum walk (CTQW) on a one-dimensional infinite lattice. The system dynamics are governed by the Hamiltonian
\begin{align}
H = -\gamma \sum_{x=-\infty}^{\infty} \left( e^{\ii\alpha}\ket{x+1}\bra{x} + e^{-\ii\alpha}\ket{x}\bra{x+1} \right),
\end{align}
where $\gamma > 0$ is the hopping rate and $\alpha$ is a phase that breaks time-reversal symmetry, leading to complex hopping amplitudes. We set $\hbar=1$ throughout. For $\gamma=1$, this model reduces to the one analyzed in Ref.~\cite{chaves2023transport}. Our work thus extends the investigation of CTQWs with complex hoppings~\cite{chaves2023transport,bhandari2023long,zimboras2013quantum,lu2016chiral,novo2021floquet,chaves2023and}.

Unlike all the works mentioned in the previous section, we define another class of tunable delocalized initial states:
\begin{align}\label{eq:initial_condition}
|\Psi(0)\rangle = \sqrt{1-D}|0\rangle + \sqrt{\frac{D}{2}}(|1\rangle + |-1\rangle).
\end{align}
The parameter $D \in [0, 1]$ controls the degree of delocalization: $D=0$ corresponds to a state localized at the origin, $|0\rangle$, while $D=1$ describes a state equally delocalized over the sites $x = \pm1$. The state defined by Eq.~\eqref{eq:initial_condition} is normalized for all $D$, as $\langle\Psi(0)|\Psi(0)\rangle = 1$.



\begin{figure}
\centering
\includegraphics[width=1\linewidth]{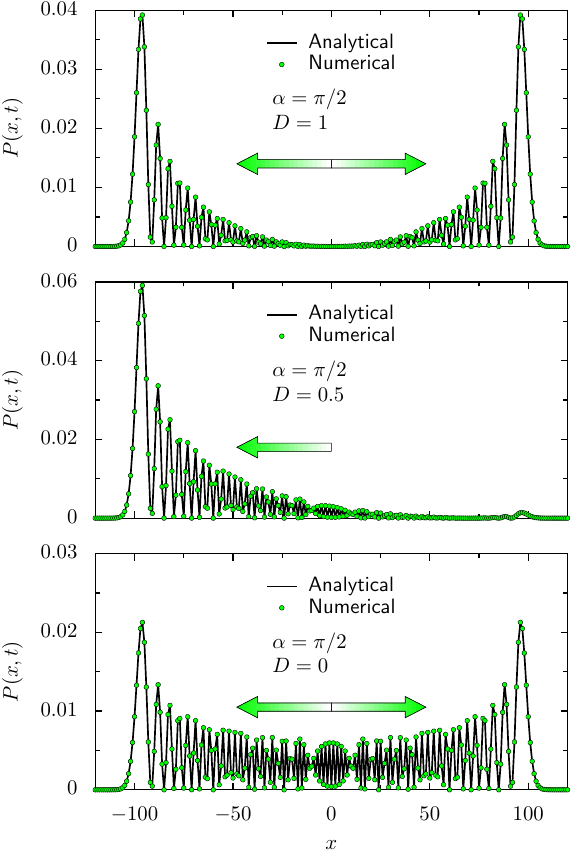}
\caption{Probability distributions for $\alpha=\pi/2$ at $t=50$. Our results show that while both localized ($D=0$, bottom) and fully delocalized ($D=1$, top) initial states yield symmetric spreading, intermediate delocalization ($D=0.5$, middle) generates a pronounced bias. 
The analytical results were obtained using Eq.~\eqref{eq:wave_function_main} and $P(x,t) = \left| \psi(x,t) \right|^2$.} 
\label{fig:prob}
\end{figure}

\begin{figure}
\centering
\includegraphics[width=1\linewidth]{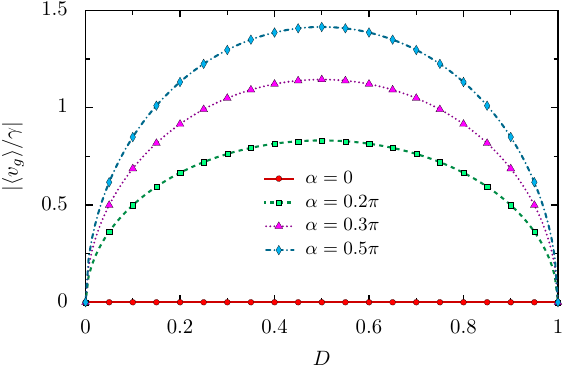}
\caption{Absolute value of the average group velocity $\abs{\langle v_g \rangle}$ [Eq.~\eqref{eq:average_main}] as a function of the delocalization parameter $D$ for various phases $\alpha$. \textcolor{black}{Solid lines represent the exact analytical results from Eq.~\eqref{eq:average_main}, while symbols show numerical simulations.} While the extreme states ($D=0$ and $D=1$) yield zero net velocity for any $\alpha$, a maximum bias emerges at intermediate delocalization ($D=0.5$), demonstrating tunable directed transport.}
\label{fig:mean_v}
\end{figure}


\paragraph{\textbf{Results}}
Due to translational symmetry, the Hamiltonian is diagonal in the momentum basis. The momentum eigenstates $|k\rangle$ are Fourier transforms of the position basis states $|x\rangle$,
\begin{align}
|k\rangle = \frac{1}{\sqrt{2\pi}} \sum_{x=-\infty}^{\infty} e^{\ii kx} |x\rangle,
\end{align}
where the quasimomentum $k$ is in the first Brillouin zone, $k \in [-\pi, \pi]$. Applying the Hamiltonian to $|k\rangle$ yields the energy dispersion relation
$E(k) = -2\gamma \cos(\alpha-k).$

In the momentum basis, the initial state is
\begin{align}
\psi(k,0) = \frac{1}{\sqrt{2\pi}} \left( \sqrt{1-D} + \sqrt{2D} \cos k \right).
\end{align}
The time-evolved wavefunction is therefore
\begin{align}
\psi(k, t) = \frac{e^{\ii 2\gamma t \cos(\alpha-k)}}{\sqrt{2\pi}} \left( \sqrt{1-D} + \sqrt{2D} \cos k \right).
\end{align}


The wavefunction in position space is obtained via the inverse Fourier transform of $\psi(k,t)$:
\begin{align}
\psi(x,t) = \frac{1}{\sqrt{2\pi}} \int_{-\pi}^{\pi} e^{\ii kx} \psi(k,t)  dk.
\end{align}
\color{black}
Using the Jacobi-Anger expansion to evaluate this integral (Sec.~I of the Supplemental Material~\cite{supp_material}), the final form of the wavefunction is found to be:
\begin{multline}\label{eq:wave_function_main}
\psi(x,t) = e^{\ii x \alpha} \bigg[ \sqrt{1-D}\, \tilde{J}_x(2\gamma t) \\
+ \sqrt{\frac{D}{2}} \left(e^{\ii\alpha}\tilde{J}_{x+1}(2\gamma t) + e^{-\ii\alpha}\tilde{J}_{x-1}(2\gamma t)\right) \bigg],
\end{multline}
\color{black}
where $\tilde{J}_n(z) \equiv \ii^n J_n(z)$ and $J_n(z)$ is the Bessel function of the first kind. The probability distribution is then given by $P(x,t) = \left| \psi(x,t) \right|^2$.

The probability distribution depends on the parameters of the initial conditions ($D$) and the Hamiltonian ($\alpha$). The nonlocal initial conditions ($D>0$) introduce two additional terms in the probability amplitude, which are modulated by a phase factor.

Figure~\ref{fig:prob} shows the analytically obtained probability distributions, which are in excellent agreement with the numerical results. For the phase value considered ($\alpha=\pi/2$), the localized ($D=0$) and fully delocalized ($D=1$) initial setups, which are commonly considered in the literature, both exhibit symmetric distributions during the system's evolution. The fully delocalized case leads to a lower probability in the central region. In contrast, the intermediate delocalization ($D=0.5$) leads to a biased spreading, despite the same phase factor value and an initially symmetric state. To further analyze this asymmetrical spreading, we calculate the average position as a function of these parameters.

\color{black}
We evaluate $\langle X\rangle(t)$ directly in momentum space using $X = \ii\partial_k$ and $\psi(k,t)=e^{-\ii E(k)t}\psi(k,0)$, which gives
\begin{align}
\langle X\rangle(t) = \int_{-\pi}^{\pi} \psi^*(k,0)\left[t\,v_g(k) + \ii\partial_k\right]\psi(k,0) dk.
\end{align}
Because $\psi(k,0)$ is real and even, the derivative term (even$\times$odd) integrates to zero, so $\langle X\rangle(t)=t\,\langle v_g\rangle$ with
\begin{align}\label{eq:average_main}
\langle v_g \rangle = -2\gamma \sin\alpha \sqrt{2D(1-D)}.
\end{align}
(See Sec.~II of the Supplemental Material~\cite{supp_material})
A net drift, characterized by $\langle v_g \rangle \neq 0$, emerges from the interplay between the initial state's delocalization ($D$) and the Hamiltonian phase ($\alpha$). The drift vanishes for the extreme cases $D=0$ and $D=1$ but is maximized for intermediate delocalization, as shown in Fig.~\ref{fig:mean_v}.
\color{black}
\begin{figure}
\centering
\includegraphics[width=1\linewidth]{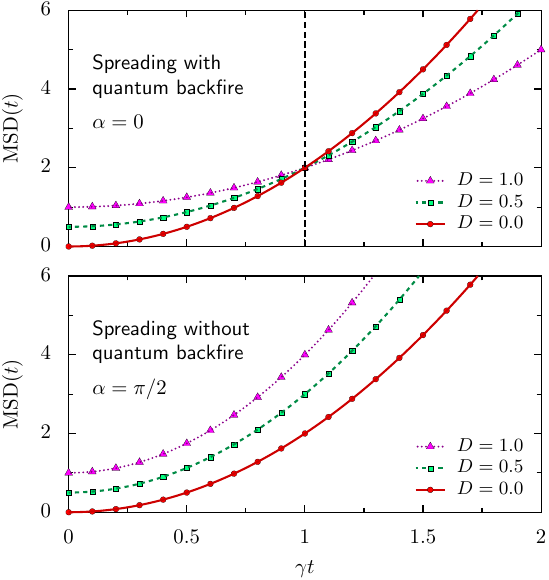}
\caption{
Time evolution of the mean square displacement (MSD) [Eq.~\eqref{eq:MSD_main}]. \textcolor{black}{Solid lines represent the exact analytical results from Eq.~\eqref{eq:MSD_main}, while symbols show numerical simulations.} The top panel ($\alpha = 0$) shows the \emph{quantum backfire effect}: (i) for $t < t_{\mathrm{cross}}$, a larger initial MSD(0) promotes a larger MSD($t$); (ii) for $t > t_{\mathrm{cross}}$, this relationship inverts: a larger initial MSD(0) produces a \emph{smaller} MSD($t$). The dashed vertical line marks $t_{\mathrm{cross}}$. The bottom panel ($\alpha = \pi/2$) exhibits no-crossing behavior: The ordering of MSD curves with respect to $D$ is preserved for all time.
}
\label{fig:std_vs_time}
\end{figure}

\begin{figure}
\centering
\includegraphics[width=1\linewidth]{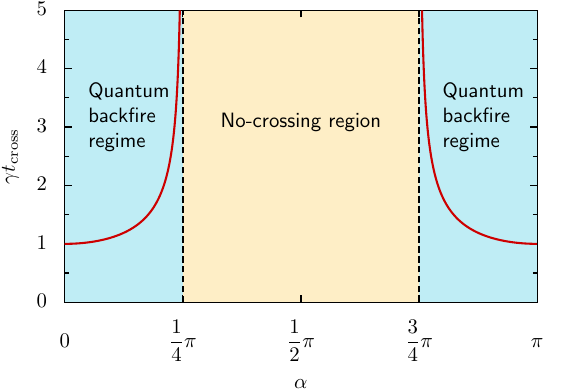}
\caption{Dependence of the MSD crossing time $t_{\mathrm{cross}}$ on the phase $\alpha$ [Eq.~\eqref{eq:t_cross}]. Two regimes are shown. 1) No-crossing (yellow, $\sin^{2}\alpha \ge 1/2$): MSD curves remain ordered for all time. 2) Crossing (blue, $\sin^{2}\alpha < 1/2$): MSD curves intersect at $t = t_{\mathrm{cross}}$. For $t > t_{\mathrm{cross}}$, the ordering inverts, demonstrating the quantum backfire effect where a greater initial delocalization ($D$) boosts short-time spreading, but is detrimental to the long-time propagation.}
\label{fig:tcross}
\end{figure}

Table~\ref{tab:main_results} summarizes the findings related to transport properties  where we see clearly that a combination between 
the initial state delocalization and the Hamiltonian phase allows a control of directional spreading.

\begin{table}[!htb]
\centering
\caption{Summary of our results for different initial conditions and properties of the Hamiltonian.  
}
\begin{tabular}{lcc}
\toprule
\textbf{Initial condition} & \textbf{Hamiltonian} & \textbf{Spreading}  \\
\midrule
 $D=0$ & Unbiased ($\alpha=n\pi$) & Unbiased  \\
 $0<D<1$ & Unbiased ($\alpha=n\pi$) & Unbiased  \\
 $D=1$ & Unbiased ($\alpha=n\pi$) & Unbiased  \\
\hline
 $D=0$ & Biased ($\alpha \neq n\pi$) & Unbiased  \\
\textbf{$0<D<1$ } 
 & 
 \textbf{Biased ($\alpha \neq n\pi$) }
  & 
\textbf{ Biased }  \\
 $D=1$ & Biased ($\alpha \neq n\pi$) & Unbiased  \\
\bottomrule
\end{tabular}
\label{tab:main_results}
\end{table}

The mean square displacement (MSD), defined as $\mathrm{MSD}(t) = \langle (X - X_0)^2 \rangle(t)$, measures the wavepacket's spreading around its initial average position $X_0 \equiv \langle X \rangle(0)$. For our initial state, symmetry dictates $\langle X \rangle(0) = 0$, simplifying the expression to $\mathrm{MSD}(t) = \langle X^2 \rangle(t)$.
\textcolor{black}{Evaluating the expectation value in the momentum basis, where $X = \ii\partial_k$, allows us to decompose the MSD into a static initial contribution and a dynamic ballistic term (see Sec.~III of the Supplemental Material for the detailed derivation~\cite{supp_material}):}
\begin{align}
\langle X^2 \rangle(t) = \langle X^2 \rangle(0) + t^2 \langle v_g^2 \rangle.
\end{align}
The initial mean square position quantifies the initial delocalization: $\langle X^2 \rangle(0) = D$. This result is intuitive, ranging from 0 for a localized state ($D=0$) to 1 for a state delocalized over sites $x = \pm1$ ($D=1$).

The time-dependent spreading is governed by the expectation value of the squared group velocity, $\langle v_g^2 \rangle$. \textcolor{black}{Substituting $v_g(k) = -2\gamma \sin(\alpha-k)$ and the initial momentum distribution into the integral $\langle v_g^2 \rangle = \int v_g(k)^2 |\psi(k,0)|^2 dk$ yields:}
\begin{align}
\langle v_g^2 \rangle = \gamma^2 \left( 2 - D + 2D\sin^2\alpha \right).
\end{align}
\textcolor{black}{The full step-by-step evaluation of this integral is provided in Sec.~III of the Supplemental Material~\cite{supp_material}.}

Combining these results yields the closed-form formula for the MSD:
\begin{align} \label{eq:MSD_main}
\mathrm{MSD}(t) = D + \gamma^2 t^2 \left( 2 - D + 2D\sin^2\alpha \right).
\end{align}
Figure~\ref{fig:std_vs_time} shows that Eq.~\eqref{eq:MSD_main} captures the full time-evolution of the wavepacket's spatial extent, from its initial value to its long-term ballistic spreading ($\sim t^2$), and highlights the non-trivial coupling between the initial condition parameter $D$ and the Hamiltonian's phase $\alpha$.

We define $t_{\mathrm{cross}}$ as the instant when the $\mathrm{MSD}(t)$ curves for different $D$ intersect, i.e., when the MSD becomes independent of the initial condition. This is obtained by setting $\partial \mathrm{MSD}/\partial D = 0$ in Eq.~\eqref{eq:MSD_main}:
\begin{equation}\label{eq:t_cross}
t_{\mathrm{cross}}(\alpha) = \frac{1}{\gamma\sqrt{\,1 - 2\sin^{2}\alpha\,}}.
\end{equation}
Thus, $t_{\mathrm{cross}}$ exists only for $1 - 2\sin^{2}\alpha > 0$ (i.e., outside the interval $\alpha \in [\pi/4,\,3\pi/4]$ modulo $\pi$) and diverges at $\alpha=\pi/4$ and $3\pi/4$.

The existence of a finite crossing time reveals a regime of counterintuitive dynamics. For a fixed time $t > t_{\mathrm{cross}}(\alpha)$ within the interval $0 < \alpha < \pi/4$, Eq. \ref{eq:MSD_main} shows that the MSD becomes a decreasing function of the delocalization parameter $D$. This means that for these times, a more delocalized initial state (higher $D$) results in a \emph{smaller} spatial spread of the wavepacket, the opposite of the intuitively expected outcome. This counterintuitive phenomenon, where the strategy of increasing initial delocalization to enhance spreading instead weakens it after a characteristic time, is formally analogous to the ``Backfire Effect'' known in cognitive science~\cite{nyhan2010when}. There, corrective evidence can paradoxically strengthen a person's misconceptions. Here, a more delocalized initial state 
improves the short-time spreading, but leads to worse asymptotic spreading performance for a proper combination of parameters. We therefore designate this as a \textbf{Quantum Backfire Effect}.

This effect is clearly visible in the post-crossing regime (blue shading in Fig. \ref{fig:tcross} for $\sin^2\alpha < 1/2$), where the MSD for $D = 1$ becomes lower than for $D = 0.5$, which in turn becomes lower than for $D = 0$. For times $t < t_{\mathrm{cross}}(\alpha)$, the intuitive ordering, where greater initial delocalization leads to greater MSD, still holds.

\begin{figure}
\centering
\includegraphics[width=1\linewidth]{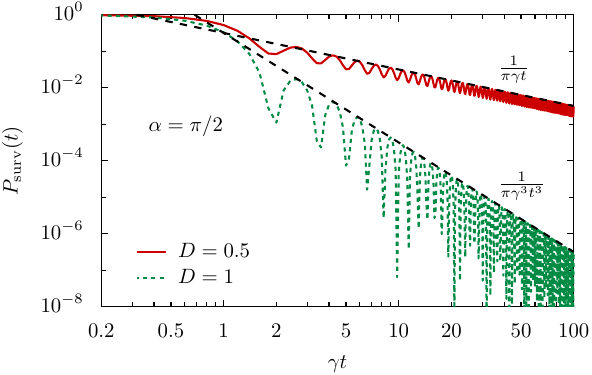}
\caption{Time evolution of $P_{\text{surv}}$ [Eq.~\eqref{eq:survival_main}] on a log-log scale for $\alpha=\pi/2$. The $D=1$ case exhibits enhanced decay ($\sim t^{-3}$), while any partially delocalized state ($D<1$) shows the standard scaling ($\sim t^{-1}$). Dashed lines show the analytical asymptotic predictions.}
\label{fig:survival_prob}
\end{figure}

The survival probability, $P_{\text{surv}}$, is an important physical quantity for classical and quantum systems~\cite{miyamoto2002various,gonulol2011survival,gonulol2013survival,krapivsky2014survival,pozzoli2021survival,segawa2023survival}.
In the context of CTQW this quantity is defined as the sum of the probabilities of finding the particle in each site of a given lattice region. In this work, we are interested in the central region of the lattice, and therefore we have:
\begin{align}
P_{\text{surv}}(t) = \sum_{x \in \{-1, 0, 1\}} P(x,t).
\end{align}
Starting with an initial condition where $ P_{\text{surv}}(0)= 1 $, the wavepacket spreading leads to a decrease in this quantity. The decay analysis of this value informs how fast the particle leaves the region.
Using the probability distribution we get \textcolor{black}{(See Sec.~IV of the Supplemental Material~\cite{supp_material})}:
\begin{align}\label{eq:survival_main}
\begin{split}
P_{\text{surv}}(t) = J_0^2(2\gamma t) &+ 2(1 - D\sin^2\alpha)J_1^2(2\gamma t)\\ + D J_2^2(2\gamma t)
& - 2D\cos(2\alpha)J_0(2\gamma t)J_2(2\gamma t).
\end{split}
\end{align}
\textcolor{black}{This exact result allows us to determine the asymptotic temporal decay by analyzing the envelope of the squared Bessel functions. For generic parameters, the survival probability follows the standard ballistic decay $P_{\text{surv}}(t) \propto t^{-1}$. However, for the fine-tuned case $D=1$ and $\alpha = \pi/2 + m\pi$ ($m \in \mathbb{Z}$), the leading-order ($t^{-1}$) terms vanish due to destructive interference. In this regime, the decay is governed by the next order in the asymptotic expansion, resulting in $P_{\text{surv}}(t) \approx 1/(\pi\gamma^3 t^3)$.}

These analytical results show that the enhanced temporal decay $t^{-3}$ occurs only in the case $D=1$ for specific values of $\alpha$. Other combinations of $D$ and $\alpha$ lead to the usual scaling with $t^{-1}$. These findings are numerically confirmed in Figure~\ref{fig:survival_prob}.

\paragraph{\textbf{Final remarks}}
We presented an analytical treatment for CTQWs with a tunable delocalization parameter $D$. 
Our main results are given by Eqs.~\eqref{eq:average_main}, \eqref{eq:MSD_main}, and \eqref{eq:survival_main}, \textcolor{black}{which relate the system's transport properties to experimentally accessible observables.}
The key insights from our work are threefold. 

First, we demonstrated \textcolor{black}{the emergence of} directional spreading from unbiased initial states with intermediate delocalization ($0 < D < 1$), \textcolor{black}{resulting from} the synergy between the initial state's delocalization and \textcolor{black}{the Hamiltonian's phase.}

Second, our analysis revealed a Quantum Backfire Effect: a phenomenon where increasing the initial state's delocalization enhances short-time spreading but \textcolor{black}{decreases} the long-term spatial spread after a crossing time, $t_{\mathrm{cross}}$. 

Third, by computing the exact survival probability, we elucidated that the transition from a standard scaling $\sim t^{-1}$ to an enhanced decay $\sim t^{-3}$ is a fine-tuned effect \textcolor{black}{specific to the fully extended initial state ($D=1$) with specific phases.}

\textcolor{black}{
The directional spreading and the backfire effect were not reported in related prior work~\cite{chaves2023transport}. Furthermore, we clarify that the breakdown of the standard scaling $P_{\mathrm{surv}}(t)\sim t^{-1}$ observed in~\cite{chaves2023transport} was a result of the specific initial condition chosen.
}

The control of transport dynamics in QWs is an important topic in quantum information science~\cite{mulken2011continuous,mulken2007quantum,hoyer2009faster,pires2020parrondo,janexperimental2020,hosaka2024parrondo,kadiri2024scouring,walczak2024parrondo,mittal2024,walczak2025parrondo,singh2019accelerated,ximenes2024parrondo,ximenes2025enhanced}.
\textcolor{black}{
Our theoretical findings are accessible to current quantum simulation platforms~\cite{wang2013physical}. For instance, photonic waveguide arrays allow for the engineering of complex hopping amplitudes via artificial gauge potentials~\cite{Song_2025,Lumer_2019} and the precise preparation of delocalized initial states~\cite{su2019experimental,PhysRevApplied.16.054036}, making the observation of these effects feasible.
}

While significant progress has been made in understanding CTQWs on networks, many questions remain~\cite{coutinho2024selected}. A promising avenue for future work would be to investigate how the phenomena we have observed in 1D systems translate to 2D lattices and more complex network structures~\cite{mulken2011continuous,biamonte2019complex}.

\paragraph{Acknowledgments}
MAP acknowledges financial support from CNPq (310093/2025-2).

\paragraph{Notes} 
This letter is accompanied by Supplemental Material~\cite{supp_material} and the source code repository~\cite{codes_ctqw}.

\bibliography{main.bib}

\pagebreak
\widetext
\begin{center}
\textbf{\large Supplemental Materials}
\end{center}
\setcounter{equation}{0}
\setcounter{figure}{0}
\setcounter{table}{0}
\setcounter{page}{1}
\makeatletter
\renewcommand{\theequation}{S\arabic{equation}}
\renewcommand{\thefigure}{S\arabic{figure}}
\renewcommand{\bibnumfmt}[1]{[S#1]}
\renewcommand{\citenumfont}[1]{S#1}

\tableofcontents

\section{Derivation of the Time-Evolved Wavefunction}

\subsection{Hamiltonian Diagonalization and Initial State}

We begin by deriving the energy dispersion relation and the momentum-space representation of the initial state. The Hamiltonian is given by:
\begin{equation}
H = -\gamma \sum_{x} \left( e^{\ii\alpha}\ket{x+1}\bra{x} + e^{-\ii\alpha}\ket{x}\bra{x+1} \right).
\end{equation}
Applying this operator to a momentum eigenstate $\ket{k} = (2\pi)^{-1/2} \sum_x e^{\ii k x} \ket{x}$, we obtain:
\begin{align}
H\ket{k} &= -\frac{\gamma}{\sqrt{2\pi}} \sum_x \left( e^{\ii\alpha}e^{\ii k x}\ket{x+1} + e^{-\ii\alpha}e^{\ii k x}\ket{x} \right) \nonumber \\
&= -\gamma \left( e^{\ii(\alpha-k)} + e^{-\ii(\alpha-k)} \right) \ket{k} \nonumber \\
&= -2\gamma \cos(\alpha-k) \ket{k}.
\end{align}
Thus, the energy dispersion relation is $E(k) = -2\gamma \cos(\alpha-k)$.

The initial state in position space is defined as:
\begin{equation}
\ket{\Psi(0)} = \sqrt{1-D}\ket{0} + \sqrt{\frac{D}{2}}(\ket{1} + \ket{-1}).
\end{equation}
Projecting this state onto the momentum basis, $\psi(k,0) = \braket{k}{\Psi(0)}$, yields:
\begin{align}
\psi(k,0) &= \frac{1}{\sqrt{2\pi}} \left( \sqrt{1-D} + \sqrt{\frac{D}{2}}(e^{-\ii k} + e^{\ii k}) \right) \nonumber \\
&= \frac{1}{\sqrt{2\pi}} \left( \sqrt{1-D} + \sqrt{2D} \cos k \right).
\end{align}
Consequently, the time-evolved state in momentum space is given by $\psi(k,t) = e^{-\ii E(k)t}\psi(k,0)$:
\begin{equation}
\psi(k, t) = \frac{e^{\ii 2\gamma t \cos(\alpha-k)}}{\sqrt{2\pi}} \left( \sqrt{1-D} + \sqrt{2D} \cos k \right).
\end{equation}

\subsection{Transformation to Position Space}

The wavefunction in position space is obtained via the inverse Fourier transform:
\begin{equation}
\psi(x,t) = \frac{1}{\sqrt{2\pi}} \int_{-\pi}^{\pi} e^{\ii kx} \psi(k,t) dk.
\end{equation}
Substituting the explicit form of $\psi(k,t)$, we have:
\begin{equation}
\psi(x,t) = \frac{1}{2\pi} \int_{-\pi}^{\pi} dk \, e^{\ii kx} e^{\ii 2\gamma t \cos(\alpha-k)} \left( \sqrt{1-D} + \sqrt{\frac{D}{2}}(e^{\ii k} + e^{-\ii k}) \right).
\end{equation}
To evaluate the integral involving the exponential of a cosine, we employ the Jacobi-Anger expansion:
\begin{equation}
e^{\ii z \cos \theta} = \sum_{n=-\infty}^{\infty} \ii^n J_n(z) e^{\ii n \theta},
\end{equation}
where $J_n(z)$ are the Bessel functions of the first kind. Identifying $z = 2\gamma t$ and $\theta = \alpha - k$, the expansion becomes:
\begin{equation}
e^{\ii 2\gamma t \cos(\alpha-k)} = \sum_{n=-\infty}^{\infty} \ii^n J_n(2\gamma t) e^{\ii n \alpha} e^{-\ii n k}.
\end{equation}
Substituting this series back into the integral expression for $\psi(x,t)$:
\begin{equation}
\psi(x,t) = \frac{1}{2\pi} \sum_{n=-\infty}^{\infty} \ii^n J_n(2\gamma t) e^{\ii n \alpha} \int_{-\pi}^{\pi} dk \, e^{\ii (x-n)k} \left( \sqrt{1-D} + \sqrt{\frac{D}{2}}e^{\ii k} + \sqrt{\frac{D}{2}}e^{-\ii k} \right).
\end{equation}
Using the orthogonality relation $\frac{1}{2\pi} \int_{-\pi}^{\pi} e^{\ii m k} dk = \delta_{m,0}$, the integration acts as a filter that selects specific values of $n$ for each term in the parenthesis. Specifically, the constant term $\sqrt{1-D}$ requires $n=x$; the term proportional to $e^{\ii k}$ requires $n=x+1$; and the term proportional to $e^{-\ii k}$ requires $n=x-1$.

Evaluating the sum at these non-zero contributions yields:
\begin{equation}
\psi(x,t) = \sqrt{1-D} \left[ \ii^x J_x(2\gamma t) e^{\ii x \alpha} \right] 
+ \sqrt{\frac{D}{2}} \left[ \ii^{x+1} J_{x+1}(2\gamma t) e^{\ii(x+1)\alpha} \right] 
+ \sqrt{\frac{D}{2}} \left[ \ii^{x-1} J_{x-1}(2\gamma t) e^{\ii(x-1)\alpha} \right].
\end{equation}
We can factor out the common phase $e^{\ii x \alpha}$ and group the terms. Note that $e^{\ii(x+1)\alpha} = e^{\ii x\alpha}e^{\ii\alpha}$ and $e^{\ii(x-1)\alpha} = e^{\ii x\alpha}e^{-\ii\alpha}$. Defining the phase-shifted Bessel function $\tilde{J}_n(z) \equiv \ii^n J_n(z)$, we arrive at the final analytical expression, consistent with the wavefunction given in the main text:
\begin{equation}
\psi(x,t) = e^{\ii x\alpha} \Bigg[ \sqrt{1-D}\,\tilde{J}_x(2\gamma t) + \sqrt{\frac{D}{2}} \left( e^{\ii\alpha}\tilde{J}_{x+1}(2\gamma t) + e^{-\ii\alpha}\tilde{J}_{x-1}(2\gamma t) \right) \Bigg].
\end{equation}

\begin{figure}[h]
\centering
\includegraphics[width=0.9\textwidth]{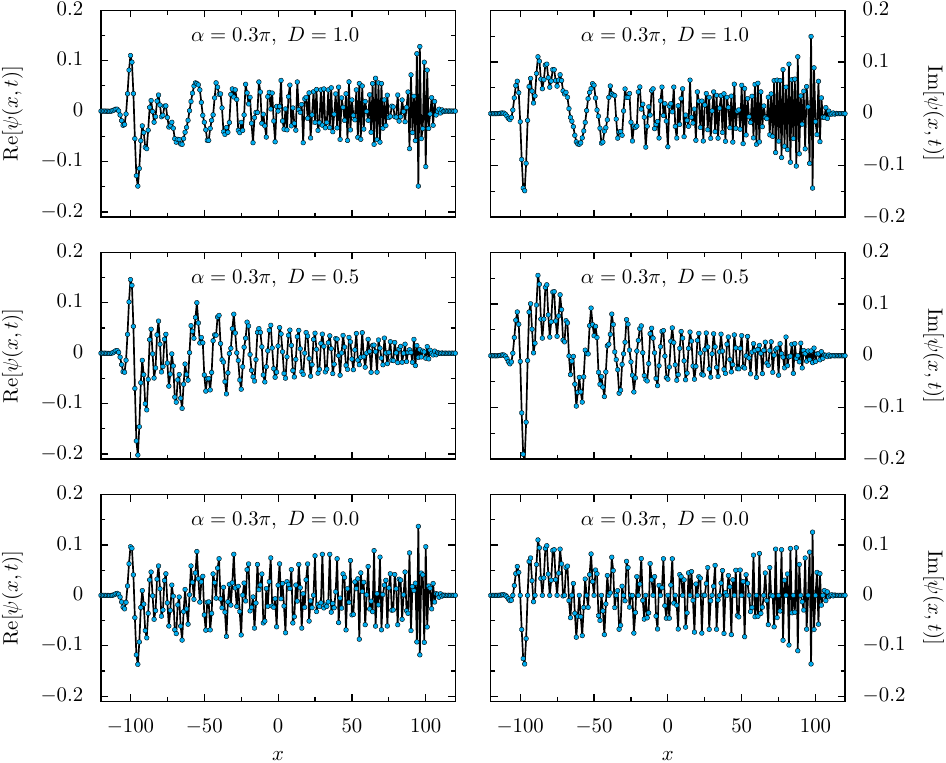}
\caption{Comparison between analytical (solid lines) and numerical (circles) results for the real and imaginary parts of the wavefunction $\psi(x,t)$ at time $\gamma t = 50$ with $\alpha = 0.3\pi$. The panels display three regimes of the delocalization parameter: localized ($D=0$, bottom), intermediate ($D=0.5$, middle), and fully delocalized ($D=1$, top). Left panels show $\Re[\psi(x,t)]$; right panels show $\Im[\psi(x,t)]$.}
\label{fig:wavefunction}
\end{figure}

Figure~\ref{fig:wavefunction} compares this analytical solution with the numerical evolution of the Schr\"odinger equation for $\gamma t = 50$ and $\alpha = 0.3\pi$. The plots for $D=0$, $D=0.5$, and $D=1$ show the real and imaginary parts of the wavefunction, demonstrating excellent agreement between the closed-form derivation and the numerical results.

\section{Average Position and Group Velocity}

In the main text, we presented the analytical result for the average position $\langle X \rangle(t)$. Here, we provide the detailed step-by-step derivation. We first derive the time dependence directly in momentum space and then explicitly evaluate the relevant integrals. Finally, we show that Ehrenfest's theorem yields the consistent result.

\subsection{Time Evolution in Momentum Space}

The expectation value of the position operator $X = \ii \frac{\partial}{\partial k}$ at time $t$ is given by:
\begin{equation}
\langle X \rangle(t) = \int_{-\pi}^{\pi} \psi^*(k,t) \left( \ii \frac{\partial}{\partial k} \right) \psi(k,t) \, dk.
\end{equation}
Substituting the time-evolved state $\psi(k,t) = e^{-\ii E(k)t}\psi(k,0)$, we apply the derivative product rule:
\begin{align}
\ii \frac{\partial \psi(k,t)}{\partial k} &= \ii \frac{\partial}{\partial k} \left[ e^{-\ii E(k)t} \psi(k,0) \right] \nonumber \\
&= \ii \left[ \left( -\ii t \frac{\partial E}{\partial k} \right) e^{-\ii E(k)t} \psi(k,0) + e^{-\ii E(k)t} \frac{\partial \psi(k,0)}{\partial k} \right] \nonumber \\
&= e^{-\ii E(k)t} \left[ t v_g(k) \psi(k,0) + \ii \frac{\partial \psi(k,0)}{\partial k} \right],
\end{align}
where we identified the group velocity function $v_g(k) = \frac{\partial E}{\partial k}$. Inserting this result back into the integral for the expectation value, we obtain:
\begin{align}
\langle X \rangle(t) &= \int_{-\pi}^{\pi} \left( e^{\ii E(k)t} \psi^*(k,0) \right) e^{-\ii E(k)t} \left[ t v_g(k) \psi(k,0) + \ii \frac{\partial \psi(k,0)}{\partial k} \right] dk \nonumber \\
&= t \int_{-\pi}^{\pi} v_g(k) |\psi(k,0)|^2 dk + \int_{-\pi}^{\pi} \psi^*(k,0) \left( \ii \frac{\partial}{\partial k} \right) \psi(k,0) dk.
\end{align}
The first term represents the drift component $t \langle v_g \rangle$. The second term is exactly the initial average position $\langle X \rangle(0)$. Thus, the motion is linear in time:
\begin{equation}
\langle X \rangle(t) = \langle X \rangle(0) + t \langle v_g \rangle.
\end{equation}

\subsection{Explicit Evaluation of Integrals}

We now calculate the two terms derived above using the specific initial state and dispersion relation of our model.

\paragraph{1. Initial Position $\langle X \rangle(0)$:}
The initial state in momentum space is real: $\psi(k,0) = \frac{1}{\sqrt{2\pi}}\left(\sqrt{1-D} + \sqrt{2D}\cos k\right)$.
Consequently, the integrand for the initial position becomes:
\begin{equation}
\psi^*(k,0) \left( \ii \frac{\partial}{\partial k} \right) \psi(k,0) = \ii \psi(k,0) \frac{\partial \psi(k,0)}{\partial k} = \frac{\ii}{2} \frac{\partial}{\partial k} \left( \psi(k,0)^2 \right).
\end{equation}
Since the wavefunction is periodic in the Brillouin zone ($k \in [-\pi, \pi]$), the integral of a total derivative vanishes. Alternatively, one can argue by parity: $\psi(k,0)$ is an even function of $k$, implying that its derivative $\frac{\partial \psi}{\partial k}$ is an odd function. The product of an even function and an odd function is odd, and its integral over a symmetric interval is zero. Therefore, $\langle X \rangle(0) = 0$.

\paragraph{2. Average Group Velocity $\langle v_g \rangle$:}
The group velocity is obtained by differentiating the dispersion relation, $v_g(k) = \frac{\partial}{\partial k} [-2\gamma \cos(\alpha-k)] = -2\gamma \sin(\alpha-k)$. We expand this using trigonometric identities as:
\begin{equation}
v_g(k) = -2\gamma (\sin\alpha \cos k - \cos\alpha \sin k).
\end{equation}
The probability distribution of the initial state is given by:
\begin{align}
|\psi(k,0)|^2 &= \frac{1}{2\pi} \left( \sqrt{1-D} + \sqrt{2D} \cos k \right)^2 \nonumber \\
&= \frac{1}{2\pi} \left[ (1-D) + 2\sqrt{2D(1-D)}\cos k + 2D\cos^2 k \right].
\end{align}
This distribution is an even function of $k$. When computing the expectation value $\langle v_g \rangle = \int v_g(k) |\psi|^2 dk$, the parity of the terms dictates the result. The term in $v_g(k)$ proportional to $\sin k$ is odd and therefore vanishes upon integration against the even probability distribution. Conversely, the term proportional to $\cos k$ is even and contributes to the integral.
Therefore, the average velocity is determined solely by the cosine component:
\begin{align}
\langle v_g \rangle &= \int_{-\pi}^{\pi} \left( -2\gamma \sin\alpha \cos k \right) |\psi(k,0)|^2 dk \nonumber \\
&= - \frac{2\gamma \sin\alpha}{2\pi} \int_{-\pi}^{\pi} \cos k \left[ (1-D) + 2\sqrt{2D(1-D)}\cos k + 2D\cos^2 k \right] dk.
\end{align}
Using the orthogonality properties of trigonometric functions, specifically that the integrals of $\cos k$ and $\cos^3 k$ vanish, while $\int \cos^2 k dk = \pi$, only the middle term of the distribution survives. This yields:
\begin{equation}
\langle v_g \rangle = - \frac{\gamma \sin\alpha}{\pi} \left[ 2\sqrt{2D(1-D)} \cdot \pi \right] = -2\gamma \sin\alpha \sqrt{2D(1-D)}.
\end{equation}
Combining these results, we obtain the final expression presented in the main text:
\begin{equation}
\langle X \rangle(t) = -2\gamma t \sin\alpha \sqrt{2D(1-D)}.
\end{equation}

\subsection{Alternative Derivation via Ehrenfest's Theorem}

For completeness, we show that Ehrenfest's theorem yields the same result immediately. For a time-independent Hamiltonian, the theorem states:
\begin{equation}
\frac{d}{dt}\langle X \rangle = \frac{1}{\ii}\langle [X, H] \rangle.
\end{equation}
In the tight-binding model, the commutator of position and Hamiltonian corresponds precisely to the group velocity operator:
\begin{equation}
\frac{1}{\ii}[X, H] = v_g(\hat{k}).
\end{equation}
Thus, $\frac{d\langle X \rangle}{dt} = \langle v_g \rangle$. Since the momentum distribution $|\psi(k,t)|^2$ is constant in time, the expectation value $\langle v_g \rangle$ is constant. Integrating with respect to time and using the initial condition $\langle X \rangle(0)=0$:
\begin{equation}
\langle X \rangle(t) = \int_0^t \langle v_g \rangle dt' = t \langle v_g \rangle.
\end{equation}
This confirms the validity of the direct derivation.



\section{Derivation of Mean Square Displacement}

The second moment is given by:
\begin{equation}
\langle X^2 \rangle(t) = \int_{-\pi}^{\pi} \psi^*(k,t) \left( -\frac{\partial^2}{\partial k^2} \right) \psi(k,t) \, dk.
\end{equation}
Using integration by parts (and noting boundary terms vanish due to periodicity), this is equivalent to:
\begin{equation}
\langle X^2 \rangle(t) = \int_{-\pi}^{\pi} \left| \frac{\partial \psi(k,t)}{\partial k} \right|^2 dk.
\end{equation}
Since $\psi(k,t) = e^{-\ii E(k)t} \psi(k,0)$, the derivative is:
\begin{align}
\frac{\partial \psi(k,t)}{\partial k} &= \left( -\ii t \frac{\partial E}{\partial k} \right) e^{-\ii E t} \psi(k,0) + e^{-\ii E t} \frac{\partial \psi(k,0)}{\partial k} \nonumber \\
&= e^{-\ii E t} \left[ -\ii t v_g(k) \psi(k,0) + \frac{\partial \psi(k,0)}{\partial k} \right].
\end{align}
Taking the squared modulus (noting that $\psi(k,0)$ and $v_g(k)$ are real functions):
\begin{equation}
\left| \frac{\partial \psi}{\partial k} \right|^2 = \left| \frac{\partial \psi(k,0)}{\partial k} - \ii t v_g(k) \psi(k,0) \right|^2.
\end{equation}
Since $\psi(k,0)$ is real, $\partial \psi(k,0)/\partial k$ is also real. The term inside is of the form $|A - \ii B|^2 = A^2 + B^2$.
\begin{equation}
\left| \frac{\partial \psi}{\partial k} \right|^2 = \left(\frac{\partial \psi(k,0)}{\partial k}\right)^2 + t^2 v_g(k)^2 \psi(k,0)^2.
\end{equation}
Integrating this gives:
\begin{equation}
\langle X^2 \rangle(t) = \int \left|\frac{\partial \psi(k,0)}{\partial k}\right|^2 dk + t^2 \int v_g(k)^2 |\psi(k,0)|^2 dk.
\end{equation}
This confirms $\langle X^2 \rangle(t) = \langle X^2 \rangle(0) + t^2 \langle v_g^2 \rangle$.

\subsection{Evaluation of $\langle v_g^2 \rangle$}

To compute the $\langle v_g^2 \rangle$, we evaluate the expectation value of the squared group velocity:
\begin{equation}
\langle v_g^2 \rangle = \int_{-\pi}^{\pi} v_g(k)^2 |\psi(k,0)|^2 dk,
\end{equation}
where $v_g(k) = -2\gamma\sin(\alpha-k)$. Using the identity $\sin^2\theta = \frac{1-\cos(2\theta)}{2}$, the squared group velocity can be written as:
\begin{equation}
v_g^2(k) = 4\gamma^2 \sin^2(\alpha-k) = 2\gamma^2 \left[ 1 - \cos(2\alpha - 2k) \right].
\end{equation}
Expanding the cosine term $\cos(2\alpha - 2k) = \cos(2\alpha)\cos(2k) + \sin(2\alpha)\sin(2k)$, the expectation value becomes:
\begin{equation}
\langle v_g^2 \rangle = 2\gamma^2 \left[ 1 - \cos(2\alpha)\langle \cos(2k) \rangle - \sin(2\alpha)\langle \sin(2k) \rangle \right].
\end{equation}
We compute the required moments using the initial probability distribution. Using $2\cos^2 k = 1 + \cos(2k)$, we expand $|\psi(k,0)|^2$ as:
\begin{align}
|\psi(k,0)|^2 &= \frac{1}{2\pi}\left( \sqrt{1-D} + \sqrt{2D}\cos k \right)^2 \nonumber \\
&= \frac{1}{2\pi}\left[ (1-D) + 2\sqrt{2D(1-D)}\cos k + 2D\cos^2 k \right] \nonumber \\
&= \frac{1}{2\pi}\left[ 1 + 2\sqrt{2D(1-D)}\cos k + D\cos(2k) \right].
\end{align}
Due to the even parity of $|\psi(k,0)|^2$, the average $\langle \sin(2k) \rangle$ vanishes. The average of $\cos(2k)$ is determined solely by the $\cos(2k)$ term in the distribution, utilizing the orthogonality relation $\int \cos(n k) \cos(m k) dk = \pi \delta_{nm}$:
\begin{equation}
\langle \cos(2k) \rangle = \int_{-\pi}^{\pi} \cos(2k) |\psi(k,0)|^2 dk = \frac{D}{2\pi} \int_{-\pi}^{\pi} \cos^2(2k) dk = \frac{D}{2}.
\end{equation}
Substituting this back into the expression for $\langle v_g^2 \rangle$:
\begin{equation}
\langle v_g^2 \rangle = \gamma^2 \left[ 2 - D\cos(2\alpha) \right] = \gamma^2 \left[ 2 - D(1 - 2\sin^2\alpha) \right] = \gamma^2 \left( 2 - D + 2D\sin^2\alpha \right).
\end{equation}
This recovers the analytical result presented in the main text.

\section{Derivation of the Survival Probability}

The survival probability is defined as the sum of probabilities within the initial subspace, $P_{\mathrm{surv}}(t) = P(0,t) + P(1,t) + P(-1,t)$. We calculate each term explicitly using the wavefunction derived in Sec. I, with the dimensionless time variable $Z=2\gamma t$.

\subsection{Probability at $x=0$}
Evaluating the wavefunction at $x=0$ yields:
\begin{equation}
\psi(0,t) = \sqrt{1-D}J_0(Z) + \sqrt{\frac{D}{2}} \left( e^{\ii\alpha}\ii J_{1}(Z) + e^{-\ii\alpha}(-\ii) J_{-1}(Z) \right).
\end{equation}
Using the identity $J_{-1}(Z) = -J_1(Z)$, this simplifies to:
\begin{align}
\psi(0,t) &= \sqrt{1-D}J_0(Z) + \ii\sqrt{\frac{D}{2}} J_1(Z) (e^{\ii\alpha} + e^{-\ii\alpha}) \nonumber \\
&= \sqrt{1-D}J_0(Z) + \ii\sqrt{2D} \cos(\alpha) J_1(Z).
\end{align}
The probability is the squared modulus:
\begin{equation}
P(0,t) = (1-D)J_0^2(Z) + 2D\cos^2(\alpha) J_1^2(Z).
\end{equation}

\subsection{Probability at $x=1$}
Evaluating the wavefunction at $x=1$, we have:
\begin{equation}
\psi(1,t) = e^{\ii\alpha} \Bigg[ \ii \sqrt{1-D} J_1(Z) + \sqrt{\frac{D}{2}} \left( -e^{\ii\alpha}J_2(Z) + e^{-\ii\alpha} J_0(Z) \right) \Bigg].
\end{equation}
Let $A = \sqrt{D/2} J_0$, $B = \ii \sqrt{1-D} J_1 e^{\ii\alpha}$, and $C = -\sqrt{D/2} J_2 e^{\ii 2\alpha}$. The probability is given by $P(1,t) = |A+B+C|^2$. Expanding and simplifying the cross terms yields:
\begin{multline}
P(1,t) = \frac{D}{2}J_0^2(Z) + (1-D)J_1^2(Z) + \frac{D}{2}J_2^2(Z) \\
- \sqrt{2D(1-D)}\sin\alpha \left[ J_0(Z)J_1(Z) - J_1(Z)J_2(Z) \right] - D\cos(2\alpha)J_0(Z)J_2(Z).
\end{multline}

\subsection{Probability at $x=-1$}
Due to the symmetry of the expansion, the expression for $x=-1$ is obtained by replacing $\alpha \to -\alpha$ in $P(1,t)$. This flips the sign of the $\sin\alpha$ term while leaving $\cos(2\alpha)$ unchanged:
\begin{multline}
P(-1,t) = \frac{D}{2}J_0^2(Z) + (1-D)J_1^2(Z) + \frac{D}{2}J_2^2(Z) \\
+ \sqrt{2D(1-D)}\sin\alpha \left[ J_0(Z)J_1(Z) - J_1(Z)J_2(Z) \right] - D\cos(2\alpha)J_0(Z)J_2(Z).
\end{multline}

\subsection{Total Survival Probability}
We obtain the total survival probability by summing the three contributions derived above. Notably, the linear cross terms proportional to $\sin\alpha$ in $P(1,t)$ and $P(-1,t)$ cancel exactly due to their opposite signs. We then group the remaining terms by their Bessel function dependence. The coefficient of the zero-order term $J_0^2$ sums to unity, since $(1-D) + D/2 + D/2 = 1$. Similarly, the coefficient of $J_2^2$ is simply $D$. The term involving $J_1^2$ has a coefficient of $2D\cos^2\alpha + 2(1-D)$, which simplifies to $2(1-D\sin^2\alpha)$. Finally, the cross term $J_0 J_2$ appears with a coefficient of $-2D\cos(2\alpha)$. Combining these results, the exact analytical expression is:
\begin{equation}
P_{\mathrm{surv}}(t) = J_0^2(Z) + 2(1-D\sin^2\alpha)J_1^2(Z) + D J_2^2(Z) - 2D\cos(2\alpha)J_0(Z)J_2(Z).
\end{equation}

\subsection{Asymptotic Analysis}

We analyze the long-time behavior of the survival probability for $Z \gg 1$ by employing the standard asymptotic expansions for Bessel functions of the first kind. For large arguments, the functions approximate to sinusoidal forms with a phase shift, given by $J_n(Z) \approx \sqrt{2/(\pi Z)} \cos(Z - n\pi/2 - \pi/4)$. Specifically, we have $J_0(Z) \approx \sqrt{2/(\pi Z)} \cos(Z-\pi/4)$, $J_1(Z) \approx \sqrt{2/(\pi Z)} \sin(Z-\pi/4)$, and $J_2(Z) \approx -\sqrt{2/(\pi Z)} \cos(Z-\pi/4)$. It is important to note that asymptotically $J_2(Z) \approx -J_0(Z)$.

Substituting these expressions into the exact formula for $P_{\mathrm{surv}}(t)$, we can organize the terms based on their oscillatory phase. The term involving $J_1^2(Z)$ contributes to a component proportional to $\sin^2(Z-\pi/4)$. The terms involving $J_0^2(Z)$, $J_2^2(Z)$, and the cross-term $J_0(Z)J_2(Z)$ all share the same phase dependence and contribute to a component proportional to $\cos^2(Z-\pi/4)$. Consequently, the asymptotic survival probability can be expressed as an envelope modulated by these oscillations:
\begin{equation}
P_{\mathrm{surv}}(t) \approx \frac{2}{\pi Z} \left[ \mathcal{C}_{\cos} \cos^2\left(Z-\frac{\pi}{4}\right) + \mathcal{C}_{\sin} \sin^2\left(Z-\frac{\pi}{4}\right) \right].
\end{equation}
The coefficient $\mathcal{C}_{\sin}$ arises solely from the $J_1^2$ term, yielding $\mathcal{C}_{\sin} = 2(1 - D\sin^2\alpha)$. The coefficient $\mathcal{C}_{\cos}$ aggregates the remaining terms; using the relation $J_0 J_2 \approx -J_0^2 \approx -J_2^2$, we obtain $\mathcal{C}_{\cos} = 1 + D - 2D\cos(2\alpha)(-1) = 1 + D + 2D\cos(2\alpha)$. The decay envelope, $P_{\mathrm{env}}(t)$, is determined by the maximum amplitude of these oscillatory terms, leading to the general scaling $P_{\mathrm{env}}(t) \approx \frac{2}{\pi Z} \max(|\mathcal{C}_{\cos}|, |\mathcal{C}_{\sin}|)$.

For generic values of the parameters $D$ and $\alpha$, at least one of the coefficients $\mathcal{C}_{\cos}$ or $\mathcal{C}_{\sin}$ remains non-zero. Since $Z = 2\gamma t$, this implies that the survival probability generally follows a power-law decay of the form $P_{\mathrm{surv}}(t) \propto t^{-1}$, corresponding to the standard ballistic spreading expected in one-dimensional quantum walks.

A distinct regime emerges if both leading-order coefficients vanish simultaneously. Setting $\mathcal{C}_{\sin} = 0$ implies $D\sin^2\alpha = 1$, which requires the physical constraints $D=1$ and $\alpha = \pi/2 + m\pi$ (with $m \in \mathbb{Z}$). Substituting these values into the cosine coefficient yields $\mathcal{C}_{\cos} = 1 + 1 + 2(-1) = 0$. In this fine-tuned scenario, the $t^{-1}$ decay channel is suppressed. To determine the dominant behavior, we return to the exact expression, which under these conditions simplifies to a perfect square: $P_{\mathrm{surv}}(t) = (J_0(Z) + J_2(Z))^2$. Utilizing the recurrence relation $J_0(Z) + J_2(Z) = \frac{2}{Z}J_1(Z)$, the survival probability becomes $P_{\mathrm{surv}}(t) = \frac{4}{Z^2} J_1^2(Z)$. Applying the asymptotic envelope for the squared Bessel function, $[J_1^2]_{\mathrm{env}} \approx \frac{2}{\pi Z}$, we find that the envelope scales as $P_{\mathrm{env}}(t) \approx \frac{8}{\pi Z^3}$. Restoring dimensions with $Z=2\gamma t$, we arrive at $P_{\mathrm{env}}(t) \approx \frac{1}{\pi \gamma^3 t^3}$, which analytically confirms the enhanced decay rate observed for the fully delocalized symmetric state under a maximally asymmetric Hamiltonian.

\section{Source code}

As a supplementary resource, we provide the detailed source code to the reader to reproduce our numerical and analytical results: \url{https://github.com/jefferson-jx/ctqw_bias_spreading_code}.

\end{document}